# Fuzzy Rule Interpolation and SNMP-MIB for Emerging Network Abnormality

Mohammad Almseidin[#], Mouhammd Al-kasassbeh[*], Szilveszter Kovacs[#]

[#] *Department of Information Technology, University of Miskolc, H-3515 Miskolc, Hungary*
*E-mail: alsaudi@iit.uni-miskolc.hu, szkovacs@iit.uni-miskolc.hu*

[*] *Computer Science Department, Princess Sumaya University for Technology, Amman, Jordan*
*E-mail: m.alkasassbeh@psut.edu.jo*

*Abstract*— **It is difficult to implement an efficient detection approach for Intrusion Detection Systems (IDS) and many factors contribute to this challenge. One such challenge concerns establishing adequate boundaries and finding a proper data source. Typical IDS detection approaches deal with raw traffics. These traffics need to be studied in depth and thoroughly investigated in order to extract the required knowledge base. Another challenge involves implementing the binary decision. This is because there are no reasonable limits between normal and attack traffics patterns. In this paper, we introduce a novel idea capable of supporting the proper data source while avoiding the issues associated with the binary decision. This paper aims to introduce a detection approach for defining abnormality by using the Fuzzy Rule Interpolation (FRI) with Simple Network Management Protocol (SNMP) Management Information Base (MIB) parameters. The strength of the proposed detection approach is based on adapting the SNMP-MIB parameters with the FRI. This proposed method eliminates the raw traffic processing component which is time consuming and requires extensive computational measures. It also eliminates the need for a complete fuzzy rule based intrusion definition. The proposed approach was tested and evaluated using an open source SNMP-MIB dataset and obtained a 93% detection rate. Additionally, when compared to other literature in which the same test-bed environment was employed along with the same number of parameters, the proposed detection approach outperformed the support vector machine and neural network. Therefore, combining the SNMP-MIB parameters with the FRI based reasoning could be beneficial for detecting intrusions, even in the case if the fuzzy rule based intrusion definition is incomplete (not fully defined).**

## I. INTRODUCTION

Nowadays, computers and network resources face different type of attacks. Modern attacks are implemented ingeniously, i.e. intruders are keeping themselves up to date with recent detection mechanisms to avoid detection. One efficient solution is the IDS detection mechanism. The IDS is currently one of the primary components used for protecting computers and networks against attacks. However, implementing an efficient IDS detection mechanism is not a straightforward task. IDS uses data sources to detect the different types of attacks. Therefore, the IDS detection mechanism requires a proper data source to be able to analyze the inbound and outbound traffics and use them to detect any abnormalities.

The typical IDS detection mechanism uses the raw network traffic as a data source to detect abnormalities within the network. Dealing with raw traffics requires in depth analysis and review to extract the information relevant for helping the attack detection [1]. The SNMP-MIB parameters can also offer the required information, yet reducing the extensive processing time necessary for analyzing the raw traffics. The SNMP-MIB can be considered as a rich data collection fetched from a series of devices for producing realistic information about the health of a network, which can be also beneficial for detecting attacks. Moreover, the binary decision related to the attack events poses another challenge for implementing an efficient IDS detection mechanism, because there is no clear decision line between normal and abnormal traffic patterns [2]. The fuzzy system provides an effective solution for dealing with the issues associated with establishing normal and abnormal decision boundaries. The fuzzy system has the ability to offer results in an explicit scheme and, consequentially, to determine the level of the attack. When as a binary decision, the degree of attack level generates an alarm, the provided attack level can help the administrator better understand the current network security status. On the other hand applying classical fuzzy reasoning methods i.e. Mamdani [3] and Takagi-Sugeno [4], the fuzzy rule base representing the attack level needs to be complete. Thus the rule base size grows exponentially with the number of the observed network parameters. In the case of partially defined incomplete fuzzy rule bases, the classical fuzzy reasoning methods could not offer the expected results for all the possible network parameter observations [5], [6].

In application areas such as IDS, it is challenging to generate complete fuzzy rule base capable of handling all possible expected observations. As a result, it is imperative to implement a fuzzy concept, especially for the IDS application area, that benefits from extending the binary decision to the continuous space and at the same time can efficiently handle the situation of the incompletely defined fuzzy rule base. Hereby, this paper proposes a novel IDS concept which implements the fuzzy rule interpolation as a detection mechanism based on the SNMP-MIB parameters. The strength of FRI methods comes from the combination of the fuzzy concept and the interpolation techniques. The FRI methods offer the required conclusion (the approximated level of attack) even when the fuzzy rules describing the attack situations are not completely defined. Consequently, the fuzzy rule base construction can be dramatically simplified. The FRI based IDS methods can achieve a satisfactory detection rate even using only a relatively small number of fuzzy rules. In this paper, we break down the implementation of the proposed FRI IDS detection approach into three main steps:

- To identify how the SNMP-MIB parameters can be used as a useful data source for detecting abnormality.
- To implement the proposed detection approach based on the strength of the fuzzy rule interpolation and the SNMP-MIB parameters.
- To highlight and discuss the difference between the proposed detection approach and other approaches that detect intrusions based on SNMP-MIB parameters.

The rest of the paper is organized as follows: section (II) illustrates recent works related to the application of intrusion detection based on SNMP-MIB parameters. Section (III), investigates and analyzes the SNMP-MIB dataset which is illustrated in detail. Then, section (IV) introduces the fuzzy rule interpolation. Section (V) introduces the proposed detection approach in detail followed by the simulation and results in section (VI). Lastly, section (VII) concludes the paper.

## II. INTRUSION DETECTION SYSTEM BASED ON SNMP-MIB

This section presents some relevant works related to the application of the detection mechanism for intrusion detection. It also provides a brief overview of different methods and approaches that are used for intrusion detection using the SNMP-MIB parameters. Typically, the SNMP is used to collect information from different data sources such as switches, routers, etc. This information is used to manage and troubleshoot different network devices. The typical IDS detection mechanism uses the raw traffic to assess the threats within connected devices. Raw traffic requires extensive investigative pre-processing to extract the required information. This investigative pre-processing is a time-consuming task for the IDS detection mechanism [1]. Therefore, the SNMP-MIB parameters offer a solution that provides the required wealth of information without needing to extensively investigate and pre-process a large amount of raw traffic.

The work of Cabrera et al. presented in [7] is one of the first attempts to apply the SNMP-MIB parameters as the data source for the IDS detection mechanism. The authors proposed a detection mechanism for Distributed Denial of Service (DDOS). Three types (Targa3, UDP Flood ad Ping Flood) of attacks were detected using the proposed detection mechanism. Altogether 90 SNMP-MIB parameters were used to detect the previous types of DDOS attacks. The SNMP-MIB parameters were collected from the simulated test-bed environment. The proposed detection mechanism was able to successfully detect the predefined DDOS attacks.

In [1], Yu et al. propose a lightweight IDS detection mechanism by adapting the machine learning algorithm to the SNMP-MIB parameters. The proposed detection mechanism avoided the raw traffic analyzation and used the statistical MIB parameters to recognize the degree of abnormality within connected devices. The authors applied the features selection algorithm to decrease a large number of SNMP-MIB parameters. The relevant parameters were determined using the correlation feature selection algorithm and the network traffic was classified using the support vector machine. Furthermore, the SNMP-MIB parameters were extracted from real-time experiments. The proposed approach achieved a fast detection time and a high rate of accuracy. The work of Yu et al. in [8] focuses on using SNMP-MIB parameters to detect flooding attacks. The authors designed the flooding attack detection mechanism by adapting the C4.5 algorithm. The SNMP-MIB parameters were collected from the simulation environment operating the flooding attack. Then, the C4.5 algorithm starts detecting and classifying the traffics based on the recorded SNMP-MIB parameters. The proposed approach was able to obtain a 93.0% detection rate.

In [9], Hsiao et al. proposed a detection mechanism for an ARP spoofing attack. The SNMP-MIB parameters were used instead of the raw traffic. The proposed detection approach was contracted into three parts. The first part was adapted to the Naive Bayesian algorithm. The second part applied the support vector machine and last part applied the C4.5 algorithm. The authors recorded their findings and highlighted both the weak and strong points for each of these parts that were used for detecting the ARP spoofing attack based on SNMP-MIB parameters. Typical performance metrics such as accuracy rate, false positive rate and missing rate were recorded for the implemented algorithms. The implemented experiments demonstrated that the C4.5 achieved the highest accuracy rate. The lowest value of false alarms was recorded by the support vector machine algorithm. The Naive Bayesian algorithm had the lowest accuracy rate within the implemented experiments.

From another perspective, the decentralized detection mechanism based on the clustering algorithm and SNMP-MIB parameters was proposed by Cerroni et al. in [10]. The proposed decentralized mechanisms were divided into the monitoring phase and the traffic detection phase. In the monitoring phase, the SNMP-MIB parameters were gathered from several agents. These parameters were forwarded to the distributed data mining algorithms for the sake of classifying the observation as either normal or abnormal. The proposed decentralized detection mechanism was tested and evaluated using the SNMP-MIB dataset and was able to detect the plausible intrusions within the dataset that related to the decentralized detection mechanisms.

In [11], Cerroni et al. introduced a new distributed data mining method in order to detect the intrusion based on SNMP-MIB parameters. The proposed method has been tested for decentralized testbed environments. The SNMP-MIB parameters were collected from the simulated network environment. Fourteen SNMP-MIB parameters were used to detect the specific type of DDOS attack. These parameters related to the IP and TCP groups. The experiments conducted reflect that the proposed mechanism obtained an acceptable detection rate.

Some other works were used in a hybrid approach, in conjunction with the SNMP-MIB parameters, to detect for abnormalities within the network traffic. In [12], Namvarasl and Ahmadzadeh proposed a hybrid approach to detect DDOS based on SNMP-MIB parameters. The proposed approach consisted of three modules; the first module was constructed for the features selection of the SNMP-MIB parameters. In the second module, the detection mechanism was generated based on high ranked SNMP-MIB parameters and the C4.5 and RIPPER were implemented to detect the intrusions. The proposed approach was tested and evaluated based on the SNMP-MIB dataset. The imported dataset consisted of 66 SNMP-MIB parameters. It also had the following type of attacks: UDP flood attack, ICMP flood attack and TCP-SYN flood attack. The proposed approach was able to detect different types of attacks within the imported SNMP-MIB dataset.

The previous works provided plausible contributions and, at the same time, supported the idea that the SNMP-MIB parameters are instrumental in detecting and recognizing the intrusion. Using SNMP-MIB parameters avoids the need for the time-consuming analysis of massive amounts of raw traffics. Previous works also shared common issues such as the difficulties associated with the detection mechanisms which suffered from a lack of clear boundaries for distinguishing between normal and abnormal traffic. Furthermore, the previous detection mechanisms did not determine the level of degree of abnormality; they only applied a binary decision to recognize the normal and abnormal traffic. In response to these issues, this paper aims to introduce a novel approach for detecting and preventing abnormalities by implementing the FRI approach with the SNMP-MIB parameters. The FRI approaches are implemented primarily to avoid binary decisions, and instead, establishing a gradient scale for distinguishing the normal and abnormal traffics. Additionally, they generate results (Detection Decision) in a clear and understandable form. Contrary to the classical fuzzy systems, the FRI approaches do not require a large amount of fuzzy rules (i.e. expert knowledge this case) for determining the level, or degree of abnormality within the protected network. Finally, the FRI approaches can produce results, even in an incompletely defined knowledge representation (fuzzy rule base).

## III. SNMP-MIB DATASET

A network attack [13] is any process used to perform malicious actions against any host inside a network with the intention of compromising the security of that network. In [14], [15], [16], a Mobile Agent (MA) was used to read the SNMP-MIB data from the local nodes that use the MIBs to store that traffic data locally. The MA was used to overcome the limitations of the centralized management system, or the IDSs. The above-mentioned works reflected that the statistical methods, based on Wiener filter and MA technology, could be joined to detect network intrusions. The suggested model intended to detect all the attacks. The MIB variables were chosen from the IF and IP groups, and the studied scenarios were the decoy port-scan, the buffer overflow, the brute force attack and the null session attack.

In this paper, the dataset we used was the same as it was introduced in [17]. This dataset was originally generated to target DoS attacks. A DoS attack is blocking legitimate user requests for services the server can provide. Such attacks can be carried out by flooding the chosen server with a high volume of traffic, thereby consuming all of the server's resources and, consequentially, preventing the server from responding to genuine requests. These attacks can be generated either from a local or remote node in a different network. DoS attacks are usually difficult to assess and prevent [18], making them one of the most challenging type threats. An even more severe threat is the DDoS attack, which is a type of flooding attack that is generated from various nodes simultaneously [19].

In this work, seven classic DoS flooding attacks are studied. The first one is the TCP-SYN attack. This attack abuses the susceptibility of the three-way handshake mechanism (SYN, SYN-ACK and ACK) operating between the host and the server when establishing the TCP/IP protocol connection. During the process, the attacker sends a SYN control packet. The server on the other side responds to the SYN request by sending an SYN-ACK packet. Meanwhile, the server stores and reserves all the resources and waits for an ACK from the sender. While the server is waiting for the ACK packet, the request remains in the memory stack. The server will not receive ACK packets from the attacker, and the attacker will send more SYN requests within a short amount time to exhaust the server's resources until it is unable to respond to any new requests [20].

The second attack is the UDP flood attack. This type of attack sends UDP packets to random ports on the victim server. When the server deals with these packets and discovers that the packets are empty, the server will then send back an error message through ICMP protocol to the sender. The server's resources, such as bandwidth which is very important for the performance of the network will be exhausted by the volume of useless or empty packets, and therefore will not be able to respond to any other requests. The UDP flood attack is typically very effective in smaller networks [21].

The third attack is the ICMP-ECHO attack. This type of attack floods the victim's bandwidth thus preventing new connections from being initiated. The PING command is used to test whether, or not the host is alive on the network. When a device receives a PING request, it will automatically reply with a message informing the sender of its status. This type of attack tricks the system by crafting a large number of ICMP packets using a spoof source IP address as the victim's IP address in order to reply directly to it later. Then it sends these packets through a network broadcast address which directs numerous hosts to send their replies to the same victim's IP address at the same time. Eventually, the high volume of reply messages will overwhelm the system and exhaust the victim's resources [17].

The fourth attack is the HTTP flood attack. This attack targets a web server and consumes the victim's resources, such as memory, CPU, bandwidth, etc. The attacker sends a huge number of valid HTTP requests (GET or POST) to a web server. Typically, these requests are generated by hosts called botnets. Each one of the bots sends a large number of legal requests at once. If there is a large number of botnets, the request rate will be higher than that is usually generated by typical users. This attack may be one of the most dangerous threats because it is hard to differentiate between normal and abnormal HTTP traffic [22]. The fifth attack is the Slowloris attack, whereby the attacker sends sessions with a high load of requests by opening multiple connections to the victim server and trying to keep these connections open as long as possible. In this case, the requests are partial HTTP requests. The attack lasts until all available sockets are reserved by the HTTP requests, causing the server to freeze in response to any legitimate connection [24]. The final attack is the Slowpost attack. Similar to the previous attack, the attacker hereby sends a complete, rather than a partial, HTTP header request, including the content length field in the post message body. The data fills the message body at the rate of one byte every two minutes. At the same time, the server remains waiting for each message body to be completed, leading to a denial of services [17].

SNMP-MIB data are rich sources providing clear statistical information about the current network device status. The SNMP-MIB is a widely deployed protocol in most network devices, and available without any additional new hardware, or software investment. By reading the MIB data, some of the major challenges of the intrusion detection can be avoided. The dataset we used contains 4998 connection records. The data is distributed into eight main classes as described in Table I.

TABLE I
TRAFFIC TYPE AND NUMBER OF GENERATED RECORDS

| No. | Type of Traffic | Number of Records |
|---|---|---|
| 1 | Normal | 600 |
| 2 | TCP-SYN | 960 |
| 3 | UDP flood | 773 |
| 4 | ICMP-ECHO | 632 |
| 5 | HTTP flood | 573 |
| 6 | Slowloris | 780 |
| 7 | Slowpost | 480 |
| 8 | Brute Force | 200 |
|  | TOTAL | 4998 |

This MIB dataset has been collected from a router. The dataset has 34 MIB variables from 5 MIB groups in MIB-II. The groups are IF, IP, TCP, UDP and ICMP. The groups and their variables are listed in Table II (see appendix).

## IV. FUZZY RULE INTERPOLATION (FRI)

The term "fuzzy logic" was introduced initially by Professor Lotfi Zadeh [23]. Fuzzy logic could be a suitable reasoning method for different application areas. The rapid growth of intrusion techniques poses a challenge for binary-decision based detection approaches. The crisp set [24] is not suitable for expressing the level of attack, which could improve the expression capabilities of the modern intrusion techniques. The fuzzy system also provide the required boundary smoothing because it can also deal with approximation, not only with precise values. The fuzzy system presents the conclusion (output) in a clear and comprehensive scheme. Establishing a fuzzy system requires several prerequisites:

- Defining the universes of the expected observations (inputs) and the possible output of the fuzzy system.
- Defining the fuzzy partitions for the inputs and outputs of the fuzzy system.
- Generating the required fuzzy rules.

Typically, IDS detection mechanism recognizes the intrusion based on crisp set that grants only membership of 0, or 1. In the classical fuzzy system, the fuzzifier transforms the crisp input parameters (observations) into fuzzy sets. The output of the fuzzy system is calculated based on the fuzzified observations [2]. The classical reasoning methods, i.e. Mamdani and Takagi-Sugeno, demands a complete fuzzy rule base to generate the desired output. Therefore, the classical reasoning methods could not infer the conclusion for any observation that is not defined in the fuzzy rule base [25].

Fig.1 shows an example of the complete fuzzy rule base in a classical fuzzy system. The parameter X is considered to be the input observation which is covered by the fuzzy rule base. Conversely, Fig.2 presents a case when an observation X is exists, but not covered by any rules of the fuzzy rule base.

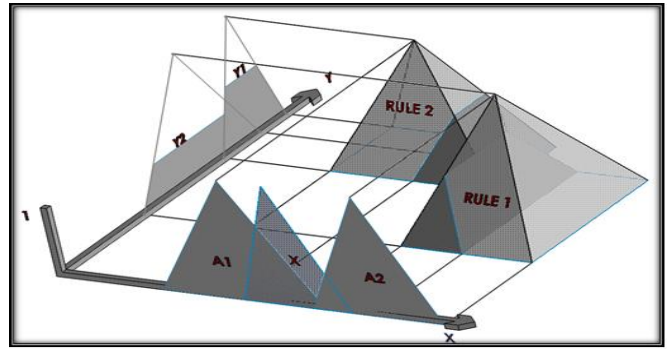

Fig.1 Complete Fuzzy Rule Base

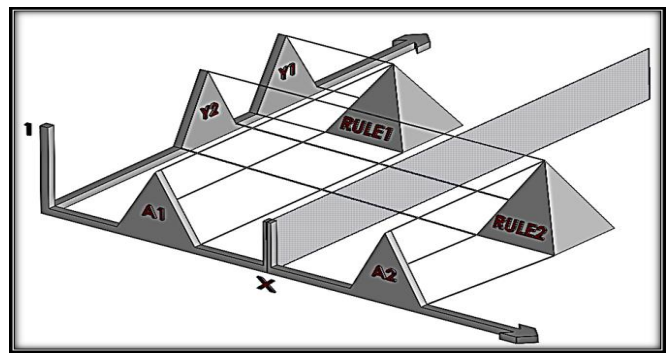

Fig.2 Incomplete Fuzzy Rule Base

In the second case (Fig.2), the classical reasoning method could not offer any conclusion; it is a method that is not always capable of satisfying the needs of some application areas (requiring conclusion for all the possible observations). In some application areas, where there is a large number of unexpected observations and the expert knowledge base cannot cover all the observation domain, it could be difficult to present the complete fuzzy rule base [2]. Therefore, the FRI methods were introduced to overcome the demand of the complete fuzzy rule base. They generate the possible inference even in cases where there is no complete fuzzy rule base. Furthermore, the FRI methods reduce the number of fuzzy rules which could be beneficial for both decreasing the complexity of the fuzzy system, and the computation time of large systems.

The proposed IDS approach is adapting the Fuzzy Interpolation based on the Vague Environment (FIVE) method as an inference engine. The FIVE method was introduced by Kovacs in [26], [27] and [35]. The concept of the vague environment, introduced by Klawon in [28], refers to the indistinguishability of a fuzzy set and a crisp value. The concept of the vague environment can be expressed by a scaling function (s). The proper scaling function (s) which describes all the fuzzy sets of a fuzzy partition, should be implemented to produce a vague environment. According to [26], [27], [28] the scaling function (s) is suitable for describing the shapes of all fuzzy set of a fuzzy partition. In the vague environment, the level of similarity between two fuzzy sets illustrates the fuzzy membership function M(x). In the vague environment, two values are ε-distinguishable if their distance is greater than ε:

$$\varepsilon > \delta_s(X_1, X_2) = \left| \int_{X_2}^{X_1} s(x)dx \right| \quad (1)$$

Likewise, $\delta_s(X_1,X_2)$ represents the vague distance for the values $X_1$ and $X_2$.

## V. THE PROPOSED DETECTION APPROACH

This section introduces the full architecture of the proposed detection approach in detail according to its main functions and prerequisites. The proposed detection approach adapts the FIVE FRI method as the inference engine. The FIVE method offers an approximated conclusion even in an incomplete fuzzy rule base.

The general structure of the proposed detection approach, as it is shown in Fig.3, is initiated by the data-cleaning stage. This stage is responsible for assembling the required information using the SNMP agents. This information is then forwarded to the SNMP manager which consists of a repository of MIB parameters. During the data cleaning stage, the MIB parameters were evaluated to determine their relevant parameters. The cleaning stage aims to reduce a large number of MIB parameters by eliminating those that are irrelevant. The detection stage of the proposed detection approach started with the Sparse Fuzzy Model Identification (SFMI), introduced by Johanyak in [29]. The outputs obtained from the data cleaning stage were passed through SFMI. It is worth mentioning that there are several operations achieved by the detection stage, including the sparse fuzzy rule generation, fuzzification and the inference engine technique.

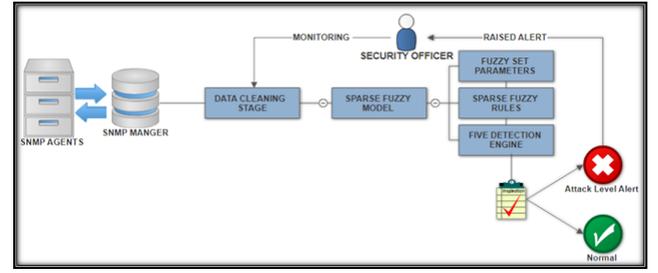

Fig.3 The General Structure of The Proposed Detection Approach

### A. The Data Cleaning Stage

The MIB parameters are characterized by the wealth of beneficial information they can offer for defining abnormality and reflect the normal and abnormal nature of the network traffics. The MIB dataset applied as an example in the rest of this paper is introduced by Al-Kasassbeh et al. in [17]. The data cleaning stage is essential for implementing the proposed detection approach. During the cleaning stage, the imported MIB dataset was divided into two parts. The first part was used for optimization, and detection approach generation, meanwhile the second part was used for the validation process. Sixty percent of the imported MIB dataset was randomly selected to train the proposed detection approach, and the rest of MIB dataset was used for the validation process.

The imported MIB dataset consists of a large number of MIB parameters. To simplify the process, the top five relevant MIB parameters [30], [17] were chosen as input parameters for detecting the abnormality in the proposed detection approach. These MIB parameters are the IP-Out-Discards and IP-In-Discards from the IP MIB group. IF-In-Discards and IF-Out-Discards from the interface MIB group and ICMP-Out-Dest-Unreachs from the ICMP MIB group. The training part is designed to generate two repositories. The first repository consists of only the intrusion traffics, and the second repository includes only the normal traffics. Algorithm 1 summarizes the data cleaning.

**Algorithm 1:** Data Cleaning Stage

**Input**: The MIB dataset
**Output**: Two repositories of normal and abnormal
1: **while** : the number of records not obtained **do**
2:    Re-sample the MIB- dataset into normal and abnormal
3:    Extract the top Five MIB parameters (ipOutDiscards, icmpOutDestUnreachs, ipInDiscardsfor, ifInDiscards and ifoutDiscards).
4:    Eliminate the rest of MIB parameters.
5:    Forward the normal traffic beside the relevant parameters to the normal repository.
6:    Forward the abnormal traffic besides the relevant parameters to the abnormal repository
7: **end while**

Consequently, 2970 instances of normal and abnormal traffics were stored in two repositories. These instances had only the top five relevant MIB parameters.

*B. The Detection Stage*

The detection stage consists of several operations including fuzzification, sparse fuzzy rule generation and adapting the inference engine. The proposed detection approach was designed and constructed using the SFMI [29]. Before constructing the proposed detection approach, the top-five relevant MIB parameters were forwarded to the SFMI.

As mentioned in section (IV), the fuzzy rule generation and fuzzy sets optimization are the necessary modelling steps for constructing the proposed detection approach. In this work, the fuzzy rule generation and the fuzzy sets optimization were adapted using the Rule Base Extension based on the Default Set Shapes (RBE-DSS) method introduced by Johanyak et al. in [31] and [32]. Algorithm 2 illustrates the sparse fuzzy rule generation process using the RBE-DSS method.

| Algorithm 2: Fuzzy Rule Generation Using RBE-DSS |
|---|
| **Input**: The training part of MIB-parameters |
| **Output**: The pool of sparse fuzzy rules |
| 1: **while** : The number of iterations not obtained **do** |
| 2:     Initiates two fuzzy rules which covered the output universe. |
| 3:     Starts modifying the initial parameters of the fuzzy set. |
| 4:     The performance parameter; Root Mean Squared Error (RMSE) of the fuzzy system computed with the modified fuzzy set parameters. |
| 5:     The fuzzy system should have the best value of RMSE for the next iteration. |
| 6:     **IF** the RMSE parameter has the same value for several iterations or the performance of the fuzzy system interrupted **Then** |
| 7:     The new fuzzy rule is generated to offer more chances of tuning. **EndIF** |
| 8:     The new fuzzy rule takes a position where the difference between the expected conclusion and the approximated conclusion is the maximum. |
| 9: **End While** |

The output from Algorithm 2 is a pool of fuzzy rules forming a sparse fuzzy rule base. Therefore, these rules could not be implemented on classical reasoning methods, such as Mamdani and Takagi Sugeno, as they are demanding complete fuzzy rule base. The FRI methods effectively reduce the total number of fuzzy rules, having 245 fuzzy rules for the FIVE method to detect the abnormality based on five MIB parameters. Table III (see appendix) presents a sample of the sparse fuzzy rule base that was generated by the RBE-DSS method.

In the RBE-DSS method, the trapezoidal membership functions were chosen to apply during the fuzzy set parameters optimization. The ICMP-Out-Dest-Unreachs, IF-Out-Discards and IF-In-Discards MIB parameters have three membership functions. These membership functions are classified into the following linguistic terms: Low, Medium and Large. The ip-In-Discards MIB parameter has four membership functions classified into the following linguistic terms: Very Low, Low, Medium and Large. Finally, the IP-Out-Discards MIB parameter represents five membership functions which are classified into the following linguistic terms: Very Low, Low, Medium, Large and Very Large. The inference engine of the proposed detection approach was performed by the FIVE method. The FIVE FRI method and SFMI source codes can be downloaded through [32].

The RBE-DSS method optimises the values of fuzzy set parameters to the maximum performance of the fuzzy IDS. Fig.4 presents the proposed detection approach's antecedent partitions•. The proposed detection approach could offer the conclusion (detection result) even in situations where some MIB parameters are not explicitly defined in the generated fuzzy rule base. Table IV (see appendix) lists the fuzzy set values optimized by the RBE-DSS method.

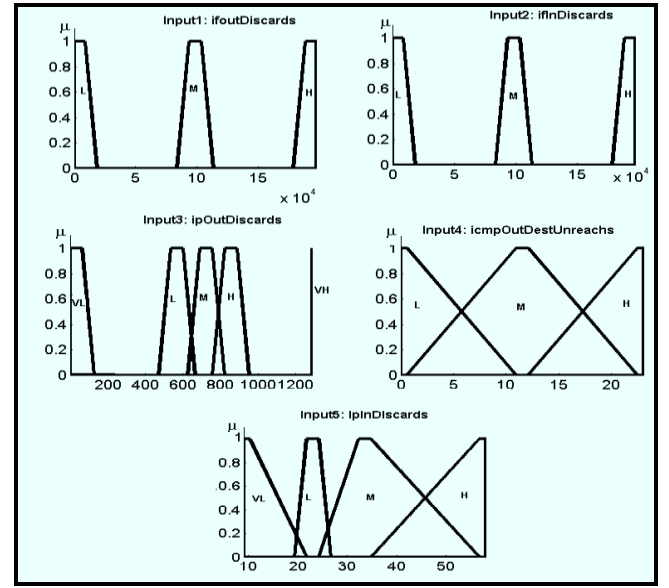

Fig. 4 The Antecedents Partitions of The Proposed Detection Approach

VI. SIMULATION AND RESULTS

This section introduces the simulation and discusses the results of the proposed detection approach in details. The implemented experiments were performed using Matlab [33] and the Fuzzy Rule Interpolation Toolbox (FRIT) [32]. The FIVE method was chosen as the inference engine of the proposed detection approach. As detailed in section (V), the total number of training data consisted of 2998 instances of normal and abnormal traffics. The training data were used to construct and optimize the proposed detection approach. The rest of MIB dataset consisted of 1998 instances which were used for the validation process. It is worth mentioning that, every observation within the SNMP-MIB dataset was presented as a fuzzy singleton. The proposed detection approach was able to generate intelligible results due to it's fuzzy nature, subsequently allowing the degree of abnormality to be determined.

Fig.5 presents the output response of the proposed detection approach in the case of abnormal instance with the parameters which are listed in Table V. The data conclude that the degree of abnormality has been determined. Subsequently, the results, which are now more concise, serve to help administrators better understanding the network's current security status.

---

• NOTE: VL = VERY LOW, L = LOW, M = MEDIUM, H = HIGH, VH= VERY HIGH.

TABLE V
ABNORAML MIB PARAMETERS EXAMPLE

| MIB Parameters | Value |
|---|---|
| IF-Out-Discards | 7270 |
| IF-In-Discards | 7270 |
| IP-Out-Discards | 1287 |
| IP-In-Discards | 9 |
| ICMP-Out-Dest-Unreachs | 0 |

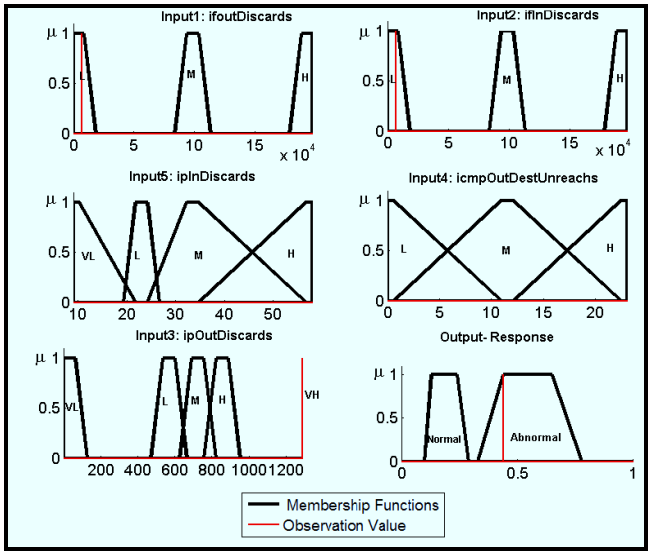

Fig.5 The Output Response of The Proposed Detection Approach

The proposed detection approach was evaluated in a two-phase process. The first phase evaluated the normal repository and the second phase was evaluated the abnormal repository. A total of 1998 MIB parameter instances were tested and evaluated. Fig.6 displays the results from both phases (normal and abnormal) of the detection approach's evaluation process.

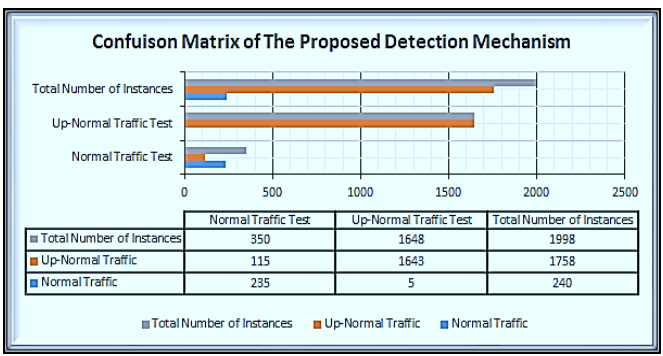

Fig.6 The Confusion Matrix of The Evaluation Process

It was concluded that five instances of normal traffics were inferred incorrectly and 115 instances of abnormal traffic were inferred incorrectly. The obtained results have been carefully analyzed and investigated to highlight the strengths of the proposed detection approach. Typically, the IDS detection mechanisms' performance depends on a few performance metrics parameters (i.e. detection rate, true positive rate, etc.) [34], which help to generate some detailed results, which can be compared with other literature results.

The performance metric formulas of [1] and [34] were used to assess the performance of the proposed fuzzy IDS detection approach. These formulas are the overall detection rate, the false negative rate, the false positive rate, the true positive rate (sensitivity), and the true negative rate (specificity). The overall detection rate (accuracy) indicates the accuracy rate for the proposed detection approach in normal and abnormal test environments. The false positive rate is an indicator of the total number of normal traffics that are recognized as abnormal (false alarm). The false negative rate is an indicator of the total number of abnormal traffics that are recognized incorrectly during the evaluation process (the total number of intrusions that successfully passed through the detection approach). Table VI presents the performance metrics for the proposed detection approach.

TABLE VI
THE PERFORMANCE METRICS FOR THE PROPOSED APPROACH

| Performance Parameter | Value | Formula |
|---|---|---|
| Sensitivity | 0.9346 | $TPR = TP / (TP + FN)$ |
| Specificity | 0.9792 | $SPC = TN / (FP + TN)$ |
| Precision | 0.9970 | $PPV = TP / (TP + FP)$ |
| False Positive Rate | 0.0208 | $FPR = FP / (FP + TN)$ |
| False Negative Rate | 0.0654 | $FNR = FN / (FN + TP)$ |
| Accuracy | 0.9399 | $ACC = (TP + TN) / (P + N)$ |

To summarize the aforementioned results, the performance of the proposed detection approach achieved satisfactory values and, at the same time, supports the idea that implementing the fuzzy rule interpolation methods for the reasoning part together with the SNMP-MIB parameters could be a promising approach in the IDS application area. Moreover, the results obtained from the proposed detection approach were compared with other literature results [30] in which the same MIB dataset and same number of relevant MIB parameters (top-5) were applied in combination with neural network, support vector machine and Bayesian network algorithms. Fig.7 compares the results between the proposed detection approach and other algorithms (neural network, support vector machine and Bayesian network).

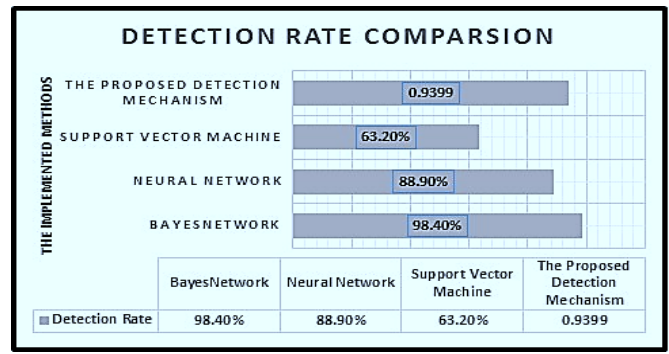

Fig.7 The Detection Rate Comparsion Results

Consequently, the implemented experiments demonstrated that the proposed detection approach achieved an acceptable accuracy rate. Moreover, it effectively reduced the false positive rate parameter. The conventional detection approaches focus on adapting the typical data mining algorithms, or the classical fuzzy reasoning methods, to be used with raw network traffics. Although FRI methods have been implemented in the IDS application area, these methods are still under investigation. Nevertheless, current research has yielded satisfactory results. The strength of FRI methods is derived from the combination of the fuzzy concept and the interpolation techniques. Therefore, the FRI methods could pose an effective solution for the boundary problem and could also handle the deficiencies of the knowledge-base representation. The proposed detection approach can be characterized by the following key points:

- The proposed FRI IDS detection approach, based on the (FIVE) FRI method, effectively smoothes the boundaries between normal and abnormal parameters by eliminating the binary decision.
- The proposed FRI IDS detection approach provides the approximated degree of abnormality even if the fuzzy rule base (knowledge representation) is not fully defined (sparse).
- The proposed FRI IDS detection approach generates comprehensive results by providing the degree of abnormality.
- The proposed FRI IDS detection approach effectively reduces the false positive rate and has obtained an acceptable detection rate value.
- The data show that the proposed FRI IDS detection approach outperformed the support vector machine and neural network algorithms in case of the example MIB parameters dataset, when it is compared to other literature results, having same conditions, in the same test-bed environment and with the same number of MIB parameters.
- The strength of the proposed detection approach is based on combining the MIB parameters with the fuzzy rule interpolation reasoning method. Thus, there is no need to deal with raw traffics which are time consuming and difficult to compute. Furthermore, this method eliminates the need for the complete fuzzy rule base.

## VII. CONCLUSION

This paper has introduced a novel method to detect abnormalities by combining the Fuzzy Interpolation based on the Vague Environment (FIVE) FRI reasoning with the Management Information Base (MIB) parameters. In that respect, there is no need to deal with raw traffic processing, which is time consuming, and difficult to compute. This method also eliminates the need for creating a complete fuzzy rule base. The MIB parameters reflect the normal and abnormal nature of the network traffics. The proposed detection approach was designed and implemented using a sparse fuzzy model identification method. The Rule Base Extension using Default Set Shapes (RBE-DSS) method was used to generate the sparse fuzzy rule base. The proposed detection approach was tested and evaluated using an open source MIB parameters dataset. The conducted experiments reflect that the proposed detection approach could effectively detect the abnormal traffics within the selected SNMP-MIB parameters dataset with 93.9% accuracy.

The results of the proposed detection approach were compared with literature results, where the same MIB variables dataset was applied in precisely the same order. The results confirmed the benefits of implementing the fuzzy rule interpolation methods for the reasoning part of the detection mechanism together with the MIB parameters as traffics sources, and also demonstrated, that the proposed FRI IDS detection approach outperforms the support vector machine and neural network based detection on SNMP-MIB parameters dataset.

## ACKNOWLEDGMENT

The described article was carried out as part of the EFOP-3.6.1-16-00011 "Younger and Renewing University – Innovative Knowledge City – institutional development of the University of Miskolc aiming at intelligent specialisation" project implemented in the framework of the Szechenyi 2020 program. The realization of this project is supported by the European Union, co-financed by the European Social Fund.

APPEDINX

TABLE II
THE SNMP MIB PARAMETERS

| Interface Group | TCP group | IP Group | ICMP Group | UDP Group |
|---|---|---|---|---|
| IFInOctets | TCPOutRsts | IPInReceives | ICMPInMsgs | UDPInDatagrams |
| IFOutOctets | TCPInSegs | IPInDelivers | ICMPInDestUnreachs | UDPOutDatagrams |
| IFoutDiscards | TCPOutSegs | IPOutRequests | ICMPOutMsgs | UDPInErrors |
| IFInUcastPkts | TCPPassiveOpens | IPOutDiscards | ICMPOutDestUnreachs | UDPNoPorts |
| IFInNUcastPkts | TCPRetransSegs | IPInDiscards | ICMPInEchos | |
| IFInDiscards | TCPCurrEstab | IPForwDatagrams | ICMPOutEchoReps | |
| IFOutUcastPkts | TCPEstabResets | IPOutNoRoutes | | |
| IFOutNUcastPkts | TCPActiveOpens | IPInAddrErrors | | |

TABLE III
THE SPARSE FUZZY RULES BASED ON THE MIB PARAMETERS

| No. | IFoutDiscards | IFInDiscards | IPOutDiscards | IPInDiscards | ICMPOutDestUnreachs | Consequence |
|---|---|---|---|---|---|---|
| 1 | Low | Low | Very Low | Very Low | Low | Normal |
| 2 | Low | Low | Very Low | Very Low | Medium | Normal |
| 3 | Low | Low | Very Low | High | High | Normal |
| 4 | Low | Low | Medium | Medium | Medium | Normal |
| 5 | Low | Medium | Very Low | Very Low | Low | abnormal |
| 6 | Medium | High | Very Low | Very Low | Low | abnormal |
| 7 | High | Low | Very Low | Very Low | Low | abnormal |
| 8 | High | High | Very High | Very Low | Low | abnormal |
| 9 | Medium | Low | Very High | Very Low | Low | Normal |
| 10 | Medium | High | Very Low | High | Medium | abnormal |
| 11 | Medium | High | Medium | Medium | High | abnormal |
| 12 | Medium | Medium | Very High | Very Low | Low | Normal |
| 13 | High | Low | Medium | Medium | Low | abnormal |
| 14 | Medium | High | Very Low | Very Low | Low | abnormal |
| 15 | Low | High | Very High | Very Low | High | abnormal |

TABLE IV
THE OPTIMIZED FUZZY SET PARAMETERS

| | Very Low | Low | Medium | High | Very High |
|---|---|---|---|---|---|
| **IpOutDiscards** | [1 1 64.8 129.1] | [472.55 536.85 601.15 665.45] | [627.93 692.23 756.53 820.83] | [758.75 823.05 887.35 951.65] | [1287 1287 1287 1287] |
| **IpInDiscards** | [9 9 10.23 21.78] | [19.33 21.78 24.23 26.68] | [24.23 32.28 34.73 56.78] | [34.73 56.78 58 58] | |
| **icmpOutDestUnreachs** | | [0 0 0.58 10.93] | [0.58 10.93 12.08 22.43] | 12.08 22.43 23 23 | |
| **ifinDiscards** | | [0 0 8602.57 18434.06] | [83567 93399 103230 113062] | [178195 188027 196630 196630] | |
| **ifoutDiscards** | | [0 0 8602.57 18434.06] | [83567 93399 103230 113062] | [178195 188027.44 196630 196630] | |